\documentclass[12pt,letterpaper,AMA,STIX1COL,doublespace]{article}
\usepackage[utf8]{inputenc}
\usepackage{color}
\usepackage{tikz}
\usepackage{amsmath}
\usepackage{bm}
\usepackage{bbm}
\usepackage{multicol}
\usepackage{geometry}
\usepackage{graphicx}
\usepackage{mathtools, nccmath}
\usepackage{relsize}
\usepackage[inline, shortlabels]{enumitem}

\usepackage{graphicx}
\graphicspath{ {./images/} }

\usepackage[numbers]{natbib}

\usepackage{booktabs,caption}
\usepackage{longtable}
\usepackage{array}
\usepackage{multirow}
\usepackage{wrapfig}
\usepackage{float}
\usepackage{colortbl}
\usepackage{pdflscape}
\usepackage{tabu}
\usepackage[flushleft]{threeparttable}
\usepackage{threeparttablex}
\usepackage{makecell}
\usepackage{xcolor}

 \usepackage{siunitx}
\usepackage{tabularray}

\usepackage{tikz}
\usetikzlibrary{shapes,decorations,arrows,calc,arrows.meta,fit,positioning}

\UseTblrLibrary{booktabs}

\begin{document}
\title{Bias in mixed models when analysing longitudinal data subject to irregular observation: when should we worry about it and how can recommended visit intervals help in specifying joint models when needed?}

\author{Rose H. Garrett$^{1,2}$, Brian M. Feldman$^{2,3,4}$, Eleanor M. Pullenayegum$^{2,1,*}$}

\date{1: Dalla Lana School of Public Health, University of Toronto, 2: Child Health Evaluative Sciences, Hospital for Sick Children, 3: Division of Rheumatology, Department of Paediatrics, University of Toronto, 4:Institute of Health Policy Management and Evaluation, Dalla Lana School of Public Health, University of Toronto\\
* Eleanor Pullenayegum,  555 University Ave, Toronto, ON, M5G 1X8, Canada; eleanor.pullenayegum@sickkids.ca
}

\maketitle












\newcommand\iid{i.i.d.}
\newcommand\pN{\mathcal{N}}
\tikzset{
    -Latex,auto,node distance =1 cm and 1 cm,semithick,
    state/.style ={ellipse, draw, minimum width = 1 cm},
    point/.style = {circle, draw, inner sep=0.04cm,fill,node contents={}},
    bidirected/.style={Latex-Latex,dashed},
    el/.style = {inner sep=2pt, align=left, sloped}
    
}  
\section*{Abstract}
In longitudinal studies using routinely collected data, such as electronic health records (EHRs), patients tend to have more measurements when they are unwell; this informative observation pattern may lead to bias. While semi-parametric approaches to modelling longitudinal data subject to irregular observation are known to be sensitive to misspecification of the visit process, parametric models may provide a more robust alternative. Robustness of parametric models on the outcome alone has been assessed under the assumption that the visit intensity is independent of the time since the last visit, given the covariates and random effects. However, this assumption of a memoryless visit process may not be realistic in the context of EHR data. In a special case which includes memory embedded into the visit process, we derive an expression for the bias in parametric models for the outcome alone and use this to identify factors that lead to increasing bias. Using simulation studies, we show that this bias is often small in practice. We suggest diagnostics for identifying the specific cases when the outcome model may be susceptible to meaningful bias, and propose a novel joint model of the outcome and visit processes that can eliminate or reduce the bias. We apply these diagnostics and the joint model to a study of juvenile dermatomyositis. We recommend that future studies using EHR data avoid relying only on the outcome model and instead first evaluate its appropriateness with our proposed diagnostics, applying our proposed joint model if necessary.\\
\\
\textbf{Keywords}: longitudinal data, irregular observation, bias, joint modelling

\section{Introduction}
In longitudinal studies using routinely collected data, patients tend to have more measurements when they are unwell and this informative observation pattern has the potential to lead to bias. Parametric models offer a powerful and useful approach to modelling longitudinal data subject to irregular observation: they are necessary for Bayesian inference and while semi-parametric models are known to be sensitive to misspecification of the visit process \citep{buuvzkova2010longitudinal, buzkovalumley}, it has been shown that parametric models are quite robust to this form of misspecification \citep{mcculloch2016biased, neuhaus2018analysis, neuhaus2020}. However, this robustness of univariate parametric models has been assessed under the specific scenario where the visit intensity is memoryless; that is, the visit intensity is independent of the time since the last visit, given the covariates and random effects. Thus, it would be useful to assess this more generally, exploring scenarios where the intensity of a visit depends on the time elapsed since the last visit. In this paper, we investigate the robustness of univariate parametric models under an informative visit process that is based on the physician’s recommendation on when the patient should return, and compare the performance of the univariate approach to a novel parametric joint model. \\
\newline
In more detail, the fundamental univariate approach for parametric analysis of longitudinal data subject to informative observation times is to use maximum likelihood to fit a standard mixed effects model on the outcome alone \citep{ep2016}. Lipsitz \citep{lipsitz2002} and Farewell \citep{farewell} discuss settings where the likelihood factorizes so that the visit process can be ignored for making likelihood-based inferences about the outcome process, and thus, consistent estimates can be obtained using the univariate approach. The issue is that in real datasets, the visit process is often non-ignorable; for example, if there is a latent, subject-specific variable that affects both the outcome and visit processes. A real-world example of such a variable, given that it is unmeasured, could be socioeconomic status; lower socioeconomic status tends to be associated with worse outcomes and also less frequent visits, since it may be difficult for people with lower socioeconomic status to get time off work and/or lose pay if away from work. In these scenarios, joint modelling of the outcome and visit processes offers an alternative analytical approach that does not rely on ignorability of the visit times \citep{ep2016, neuhaus2018analysis, gasparini, ryu2007}.\\
\newline
However, Neuhaus et al \citep{neuhaus2018analysis} argue that in practice, joint models are difficult to specify correctly, since the intervals between visits are highly irregular and often depend on information that is not available to the analyst. Thus, it is pertinent to consider the consequences of opting for the simpler univariate approach, knowing that the modelling assumptions will be violated by the non-ignorable visit processes that are typically encountered in real-world datasets. McCulloch et al \citep{mcculloch2016biased} investigated this problem; they identified under what conditions a standard mixed effects model will yield biased estimates of the parameters of interest and formulated expressions for the size of the bias under different data-generating mechanisms. They found that estimates of the fixed effects may be biased if they have associated random effects, but the effects of covariates unconnected to the random effects can be consistently estimated. Neuhaus et al \citep{neuhaus2018analysis,neuhaus2020}  expanded upon this work, examining additional visit scenarios and modelling approaches, and found that the identified bias tends to be fairly small in practice, and they also provided suggestions on study design strategies for reducing this small bias even further. It is important to note that the approaches taken in this series of papers assume that the visit intensity is independent of the time since the last visit, given the covariates and random effects. \\
\newline
Although the data generating mechanisms used in the above work \citep{mcculloch2016biased, neuhaus2018analysis, neuhaus2020} are reflective of a range of real-life visit scenarios, in other contexts, such as clinic-based cohorts, it is helpful to embed memory into the visit process. When patients are advised to come back at set intervals, the time since the last visit will be a stronger predictor of visit intensity than the full length of time to the visit measured from a reference starting point (e.g., date of diagnosis, beginning of study) \citep{nazeri2016estimation}. The basic idea is that if a patient visited today, they are unlikely to come back tomorrow, and often the time at which patients visit is influenced by when they were advised to come back (e.g. physician recommends they return in 6 months).  In addition, in clinic-based cohorts, the recommendations on when to return are highly variable depending on the patient’s disease status. Thus, in this paper, we formulate the visit process based on intervals between visits, rather than the visit times themselves, and we also incorporate the close correspondence between the visit intervals and the physician’s recommendations on when to return. \\
\newline
The main objective of this paper is to generalize McCulloch et al's \citep{mcculloch2016biased} previous results by investigating the robustness of standard mixed effects models under an informative visit process generated based on intervals between visits. The secondary objective is to formulate a novel joint model of the intertwined visit and outcome processes that incorporates the physician's recommendation on when to return, which has not been used in previous modelling approaches. The performance of the proposed joint model will also be compared to the aforementioned standard mixed effects model on the disease outcome alone. \\
\newline
The paper is structured as follows: in Section 2  we develop theoretical results that parallel those of McCulloch et al\citep{mcculloch2016biased} in this more general visit process setting, in Section 3 we formulate our novel joint model, and in Section 4 we outline the design and present the results of an extensive simulation study  that evaluates the performance of our proposed joint model and the standard mixed effects model under a variety of settings. Finally, in Section 5, we illustrate our approach by analyzing a previously reported clinic-based cohort of patients diagnosed with juvenile dermatomyositis (JDM).
\section{Theory}
We begin by defining the key notation and simplifying assumptions we will use to show our theoretical results. Then we outline the basic set-up used in \citep{mcculloch2016biased} (adjusting their notation to be consistent with our paper), and review their main results. Drawing on this foundation, we derive an expression for the bias of the fixed effect estimates in the outcome model under our proposed visit process specification. \\
\newline
In addition to the general expression, we look at special cases of simplified scenarios that are easier to interpret, and we identify under what conditions we can expect to see substantial bias in practice.
\subsection{Key notation and assumptions}
We let $Y_{i}(t)$ represent the outcome measurement at time $t$ on subject $i$. Initially, we will assume that  observations are conditionally independent given the subject-specific random effects, $b_{i}$. We will also assume a normally distributed outcome, so that we can use a linear mixed model framework. In addition, we initially assume that the outcome and visit processes are conditionally independent, given $b_{i}$.\\
\newline
In terms of the visit process, we use the notation $N_{i}(t)$ to denote the counting process for the visit times for subject $i$, and we denote the visit time for subject $i$ at visit $j$ as $T_{ij}$. We further use the counting process notation $\Delta N_{i}(t)=N(t)-N(t^{-})$, where $t^{-}$ is the instant of time right before $t$. That is, $N(t^{-})=\lim_{s \uparrow t} N(s)$. If $\Delta N_{i}(t)=1$, a visit occurred at time $t$. We also use the notation $\bar{N}_{i}(t)$ to denote the history of the visit process (i.e. the history of when all the visits occurred) up to time $t$. Finally, we let $\tau$ denote the timing of the end of the follow-up period.
\subsection{McCulloch et al. (2016) set-up and results}
We will reference the McCulloch linear mixed outcome model set-up \citep{mcculloch2016biased}
\begin{align}\label{eqn:yneuhaus}
    Y_{i}(t) &=  X_{i}^{T}(t)\beta + Z_{i}^{T}(t)b_{i} + \epsilon_{i}(t),
\end{align}
where $\epsilon_{i}(t) \sim N(0, \sigma^2_{\epsilon})$, $b_{i} \sim N(0, \Sigma_{b})$ is the $q$-dimensional vector of random effects for subject $i$, $X_{i}(t)$ denotes the $p$-dimensional vector of covariates for subject $i$ at time $t$, $\beta$ is the $p$-dimensional vector of fixed effects, and $Z_{i}(t)$ represents the $q \times 1$ model matrix for the random effects.\\
\newline
The informative visit process model is formulated as
\begin{align}\label{eqn:visitprocess_neuhaus}
    P\Big(Y_{i}(t) \hspace{3 pt} \text{is observed} \mid b_{i}\Big)=P\Big(\Delta N_{i}(t)=1 \mid b_{i} \Big)=\exp{(\mu_{i}(t) + \gamma_{i}^{T}(t) b_{i})},
\end{align}
where $\mu_{i}(t)$ denotes the fixed effects portion of the model (covariates together with covariate effects), $\gamma_{i}(t)$ governs the strength and directionality of the association between the random effects and whether or not data are observed. $\mu_{i}(t)$ and $\gamma_{i}(t)$ can depend arbitrarily on either fixed or time-varying covariates. We note that if we were to discretize time over a fine grid (e.g. days) of unit length $\Delta t$, then this probability is approximately equal to $\lambda(t)\Delta t$ where $\lambda(t)$ is the visit intensity.\\
\newline
The distribution of the observed outcomes is given by $Y_{i}(t) \mid \Delta N_{i}(t)=1 \sim N\Big( X_{i}^{T}(t)\beta + Z_{i}^{T}(t) \Sigma_b \gamma_{i}(t), \hspace{3pt} \sigma^2_{\epsilon} +  Z_{i}^{T}(t)\Sigma_{b} Z_{i}(t)\Big)$ \citep{mcculloch2016biased}. This result ultimately leads to the paper's final conclusion that when using a univariate mixed model for irregularly observed longitudinal data, estimators of parameters of covariates associated with the random effects will be biased, but the other covariate effect parameters will be consistent and estimated with little or no bias. They note that there does not necessarily need to be a direct association with the random effects in the visit process to incur bias. For example, suppose the outcome model has both a random intercept and random slope, but the visit process depends only on the random intercept. In this case, the fixed effect for the slope will still be biased if the random intercept and slope are correlated. \\
\newline
Model (\ref{eqn:visitprocess_neuhaus}) is memoryless; however, in the context of routine follow-up with heterogeneous disease courses, such as lupus, the timing of the next visit depends on when the previous visit occurred. Models built around visit intervals provide an option for representing such a visit process with memory of when the previous visit occurred. In the next section, we propose an initial set-up for embedding memory into the visit process, making some simplifications to allow for derivations of theoretical results paralleling those of McCulloch et al \citep{mcculloch2016biased}.
\subsection{Embedding memory into the visit process model: the general case}
We use a simplified version of the set-up given in (\ref{eqn:yneuhaus}) for the disease outcome submodel, considering only baseline covariates, $X_{i}$ (but see Appendix for the time-varying case). For the visit process submodel, we model the interval between visit $j$ and $j+1$, which we will denote as $S_{ij}$. We will assume that at the end of each visit, the patient is told when to come back and they adhere to it perfectly. Therefore, we know the value of $S$ at the time of the previous visit. This simplifies matters as it allows us to condition on the observed history available at the previous visit, and we do not have to consider censoring of the final $S$. We initially assume $S_{ij}$ is normally distributed, noting that we will discuss the model's placement of a non-zero probability on the event that $S_{ij} < 0 $ later, and we will also later relax the normality assumption. We use the linear mixed submodel
\begin{align}\label{eqn:Smodel}
S_{ij}= H_{i}^{T} \alpha + \gamma_{i}^{T}b_{i}+ \eta_{ij}, 
\end{align}
where $\eta_{ij} \sim N(0, \sigma^2_{\eta})$, $H_{i}$ denotes the $r$-dimensional vector of baseline covariates for subject $i$, $\alpha$ is the $r$-dimensional vector of fixed effects, and $\gamma_{i}$ serves the same function as in (\ref{eqn:visitprocess_neuhaus}) except that the association is now with the size of the intervals between visits rather than whether or not a visit occurs. That is, $\gamma_{i}$ links the outcome to the visit process. We assume that $Y$ is scaled so that higher values indicate worse disease symptoms. As in \citep{mcculloch2016biased}, we assume that the outcome and visit processes are independent of each other, conditional on $b_{i}$. 

Since we have built memory into the visit process, we consider the conditional distribution of $Y_{i}(t)$ given $\bar{N}_{i}(\tau)$, which is equivalent to conditioning on the set of visit intervals up to the end of the follow-up period, denoted as $\bar{S}_i$. 

In the Appendix we show that
\begin{align}
    b_{i} \mid \bar{S_{i}} \sim MVN \bigg(\frac{ \Sigma_{*}\gamma_{i}}{\sigma^2_{\eta}}  \Big( U_{i} -N_{i}(\tau)H_{i}^{T} \alpha  \Big), \Sigma_{*}\bigg), \label{gen_nu}
\end{align}
where $U_{i}=\sum_{j=1}^{N_{i}(\tau)} S_{ij}$ and $\Sigma_{*}=  {\bigg( \frac{N_{i}(\tau)\big(\gamma_{i} \gamma_{i}^{T}\big)
}{\sigma^2_{\eta}} + \Sigma_{b}^{-1} \bigg)}^{-1}$. \\
\newline
In more detail, $U_{i}$ is the sum of the intervals between visits. We note that $U_i > \tau$, since $U_{i}$ includes the interval between the final visit recorded in the study period and the next visit that occurs after the study ends ($S_{iN_{i}(\tau)}$). In addition, $H_{i}^{T} \alpha $ represents the average interval length, and thus, $N_{i}(\tau)H_{i}^{T}\alpha$ is the expected sum of the intervals between visits. \\
\newline
We use the general expression for the maximum likelihood estimator (MLE) for the fixed effects in a linear mixed model
\begin{align}
     \hat{\beta} &= {\bigg( \sum_{i=1}^{M} {X_{i}}w_{i} X_{i}^{T} \bigg)}^{-1} \bigg( \sum_{i=1}^{M} {X_{i}}w_{i} \bar{Y}_{i} \bigg),
\end{align}
where $M$ denotes the number of individuals in the sample, $\bar{Y}_{i}= \frac{1}{N_{i}(\tau)}\sum_{j=1}^{N_{i}(\tau)} Y_{i}(T_{ij})$, and $w_{i}={\Big({Var}(\bar{Y}_{i})\Big)}^{-1}={\Big( Z_{i}^{T}\Sigma_{b}Z_{i} + \frac{\sigma^2_{\epsilon}}{N_{i}(\tau)}\Big)}^{-1}$. \\
\newline
We take $\text{bias}=E(\hat{\beta} \mid \bar{S_{i}})-\beta$, and get
\begin{align}
   \text{bias}= {\bigg( \sum_{i=1}^{M} {X_{i}}w_{i} X_{i}^{T} \bigg)}^{-1} \bigg( \sum_{i=1}^{M}  {X_{i}}w_{i} Z_{i}^{T}\frac{ \Sigma_{*}\gamma_{i}}{\sigma^2_{\eta}}  \Big( U_{i} -N_{i}(\tau)H_{i}^{T} \alpha  \Big) \bigg). \label{baseline_generalcase}
\end{align}
We note that this general expression is difficult to interpret. We simplify matters further by considering several special cases.

\subsection{Intercept-only model}
\begin{figure}
\begin{center}
  \includegraphics[scale=0.8]{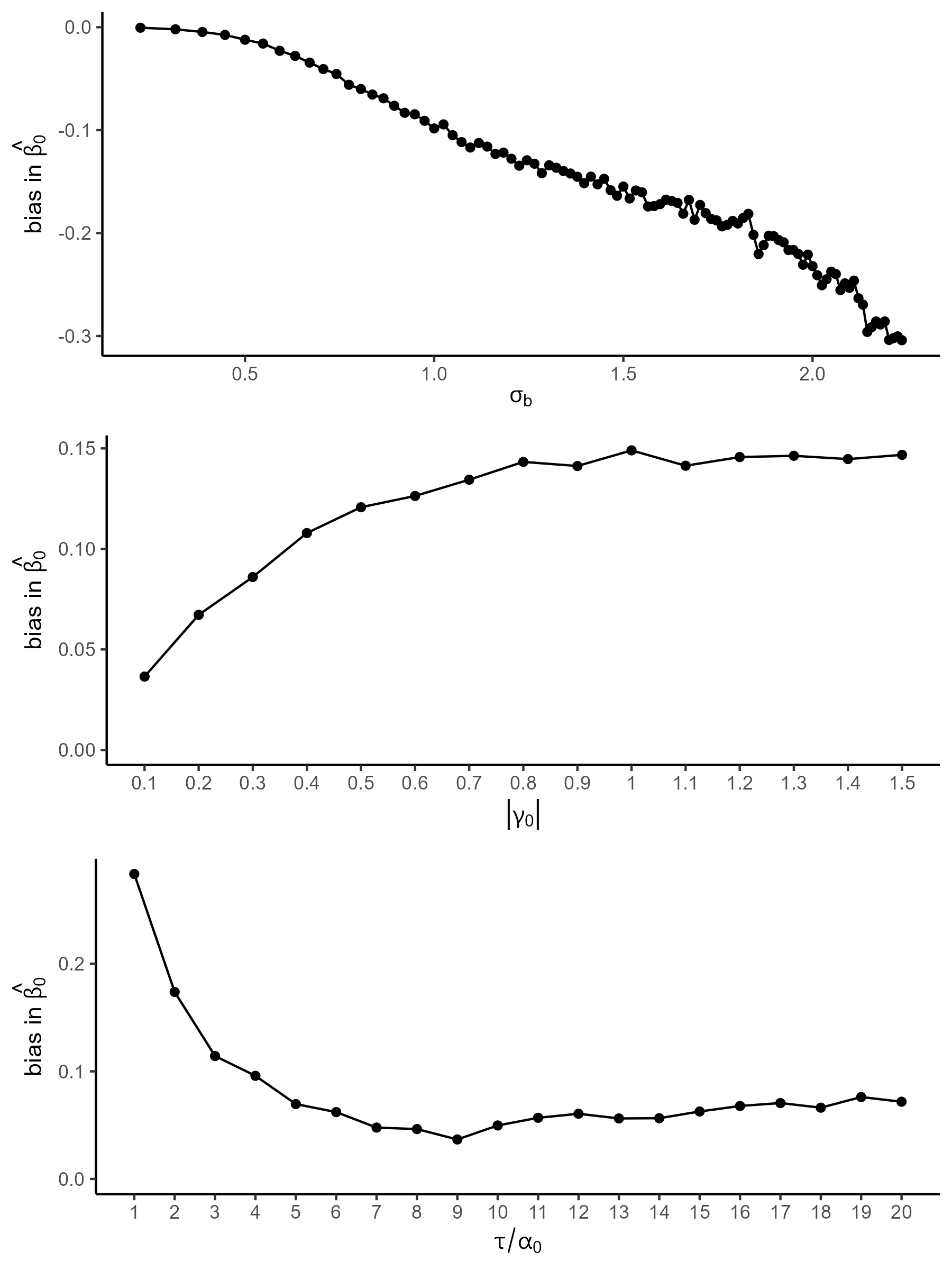}
  \caption{Trends in bias as the standard deviation of the random intercept, strength of the association between the visit and outcome models is increased, and average number of visits per individual(by decreasing the average visit interval length) is increased, respectively, holding all other parameters constant ($\sigma_{b}=\sqrt{2}$, $\sigma_{\eta}=1$, $\tau=200$, $\alpha_{0}=200/3$, $\sigma_{\epsilon}=5$ and $\gamma_{0}=-1$).} \label{fig:fourpanelplot}
  \end{center}
\end{figure}

We set $X_{i}(t)=Z_{i}(t)=1$ in Equation \eqref{eqn:yneuhaus} and $H_{i}=1$ with a constant $\gamma_{i}=\gamma_{0}$ in Equation \eqref{eqn:Smodel}. That is, we have submodels $Y_{i}(t) = \beta_0 + b_{0i}  + \epsilon_{i}(t)$ and $S_{ij} = \alpha_0 + \gamma_0b_{0i} + \eta_{ij}$, with $b_{0i}  \sim N(0, \sigma^2_{b})$.\\
\newline
This results in 
\begin{align}
    \text{bias in } \hat{\beta_{0}} = \bigg(\frac{1}{\sum_{i=1}^{M}w_{i}}\bigg)\bigg( \sum_{i=1}^{M} w_{i}\frac{\frac{1}{\gamma_{0}}\big(U_{i} - N_{i}(\tau)\alpha_0 \big)}{N_{i}(\tau) +
\frac{\sigma^{2}_{\eta}}{\sigma^{2}_{b}\gamma_{0}^{2}}}  \bigg), \label{simplebias}
\end{align}
where $w_{i}$ simplifies to ${(\sigma^2_{b} + \frac{\sigma^2_{\epsilon}}{N_i(\tau)})}^{-1}$. Using this formula, we can set up hypotheses about how the bias changes as a function of the parameters.\\
\newline
Firstly, we hypothesize that increasing the random effect variance $\sigma^2_{b}$ should increase the bias: more subject-to-subject variation in health status should lead to more variation in the visit interval size, and thus a larger disparity between the observed sum of intervals $U_{i}$ and expected sum $N_{i}(\tau) \alpha_{0}$ in Equation \eqref{simplebias}, and therefore larger bias. Secondly, we hypothesize that increasing the magnitude of $\gamma_{0}$ should also increase the bias for the intuitive reason that this would increase the strength of the association between the visit and outcome processes, and since the denominator in \eqref{simplebias} gets smaller as $\gamma_{0}$ increases. Our third hypothesis is that the bias will decrease as the average total number of visits per person, which can be approximated by $\tau / {\alpha_0} $, increases. This is because this leads to a smaller disparity between $U_{i}$ and $N_{i}(\tau) \alpha_{0}$ in \eqref{simplebias}, since as  $\tau / {\alpha_0} $ becomes large compared to the variance in $S_{ij}$ (which is $\gamma_0^2\sigma_b^2 + \sigma_{\eta}^2$), $U_i$ will get closer to $\tau$ and $N_i(\tau)\alpha_0$ will get closer to $\tau$ too, so the bias should decrease. \\
\newline
Next, we test these three hypotheses through simulation, with the overarching goal of assessing the size of the bias in practice. We simulated the intervals between visits according to the simplified intercept-only $S_{ij}$ model, by setting $\tau$, and the parameters $\alpha_0$, $\sigma_{b}$, $\gamma_{0}$, and $\sigma_{\eta}$, and then generating $S_{ij}$ values until $U_{i}$ exceeds $\tau$. We note that a small fraction of the time, Model \eqref{eqn:Smodel} generated small or negative values of $S_{ij}$, and in these relatively rare cases, we truncated the $S_{ij}$ at one week. The number of intervals ($S_{ij}$'s) generated gave us the  value of $N_{i}(\tau)$, and we also specified $\sigma_{\epsilon}$ in order to calculate the weights $w_{i}$. We then applied  \eqref{simplebias} to compute the bias (i.e. we did not fit mixed models). We set $M=100,000$ individuals. \\
\newline
Figure \ref{fig:fourpanelplot} confirms that the magnitude of the bias increases with increasing random intercept variance, increasing magnitude of correlation between the random intercept and visit intervals, and with increasing average visit interval length.

\subsection{Baseline covariates unconnected to visit process}
For this set-up, we consider baseline covariates in the $Y$ submodel that do not have a corresponding random effect, and are not connected to the visit process in any way. That is, these covariates are independent of $N_{i}(\tau)$. In more detail, we make the additional assumptions here that $X_{1i} \perp H_{i}, b_{i}, X_{2i}$ and $X_{1i}$ does not have an associated random slope. For ease of interpretation, we consider a univariate $X_{1i}$ but these results hold for the multivariate case as well. The submodels are specified as 
\begin{align}
    Y_{i}(t) &= \beta_0 + \beta_{1} X_{1i} + \beta_{2} X_{2i} +b_{0i} + b_{2i}X_{2i} + \epsilon_{i}(t) \label{Ycase2} \\
    S_{ij}&= \alpha_0 + \alpha_1 H_{i} + \gamma_{0}b_{0i} + \gamma_{2} b_{2i}X_{2i} + \eta_{ij} , \label{Scase2}
\end{align}
where we assume, without loss of generality, $X_{1i}$ is standardized to have mean 0 and variance 1.\\
\newline
We note that since under this set-up, $X_{1i} \perp W_{i}, \hspace{2 pt} U_{i},  \hspace{2 pt} N_{i}(\tau)$ in \eqref{baseline_generalcase}, $\hat{\beta_{1}}$ will be unbiased. However, $\hat{\beta_{0}}$ and $\hat{\beta_{2}
}$ will still be biased. See Appendix for the details of the theory underlying these results. \\
\newline
 Thus, while intercepts and covariate effects with random effects are estimated subject to bias, coefficients of covariates that are independent of the visit process are estimated without bias.
\subsection{Random intercept model with shared binary baseline covariate}
We follow \eqref{Ycase2} and \eqref{Scase2} but consider a single covariate $X_{i}=H_{i}$ and set $X_{i}\in \{ 0, 1 \}$.\\
\newline
Under this set-up, we have (see Appendix for derivation)
\begin{align}
  \text{bias in } \hat{\beta_{1}}=  \smashoperator{\sum_{i : x_{i}=1}} w_{i}^{*} \bigg[ \frac{\frac{1}{\gamma_{0}}\big(U_{i} - N_{i}(\tau)(\alpha_0 + \alpha_{1}) \big)}{N_{i}(\tau) +
		r} \bigg]  - \smashoperator{\sum_{i : x_{i}=0}} w_{i}^{*} \bigg[ \frac{\frac{1}{\gamma_{0}}\big(U_{i} - N_{i}(\tau)\alpha_0  \big)}{N_{i}(\tau) +
		r} \bigg],
\end{align}
where we have normalized the weights so that $ \smashoperator{\sum_{i : x_{i}=1}} w_{i}^{*}=1$ and $ \smashoperator{\sum_{i : x_{i}=0}} w_{i}^{*}=1$.\\
\newline
As $\alpha_1$ gets further away from zero and more positive, we expect the bias to increase because the gaps between visits get larger, and thus for a set $\tau$, there are fewer visits per patient. We can investigate this hypothesis through simulation. We follow the simulation procedure described in Section 2.4 under the intercept-only model, setting the additional parameter $\alpha_1$ and generating $X_{i} \sim \text{Bernoulli}(p=0.5)$. 
\begin{figure}[ht]
\begin{center}
  \includegraphics{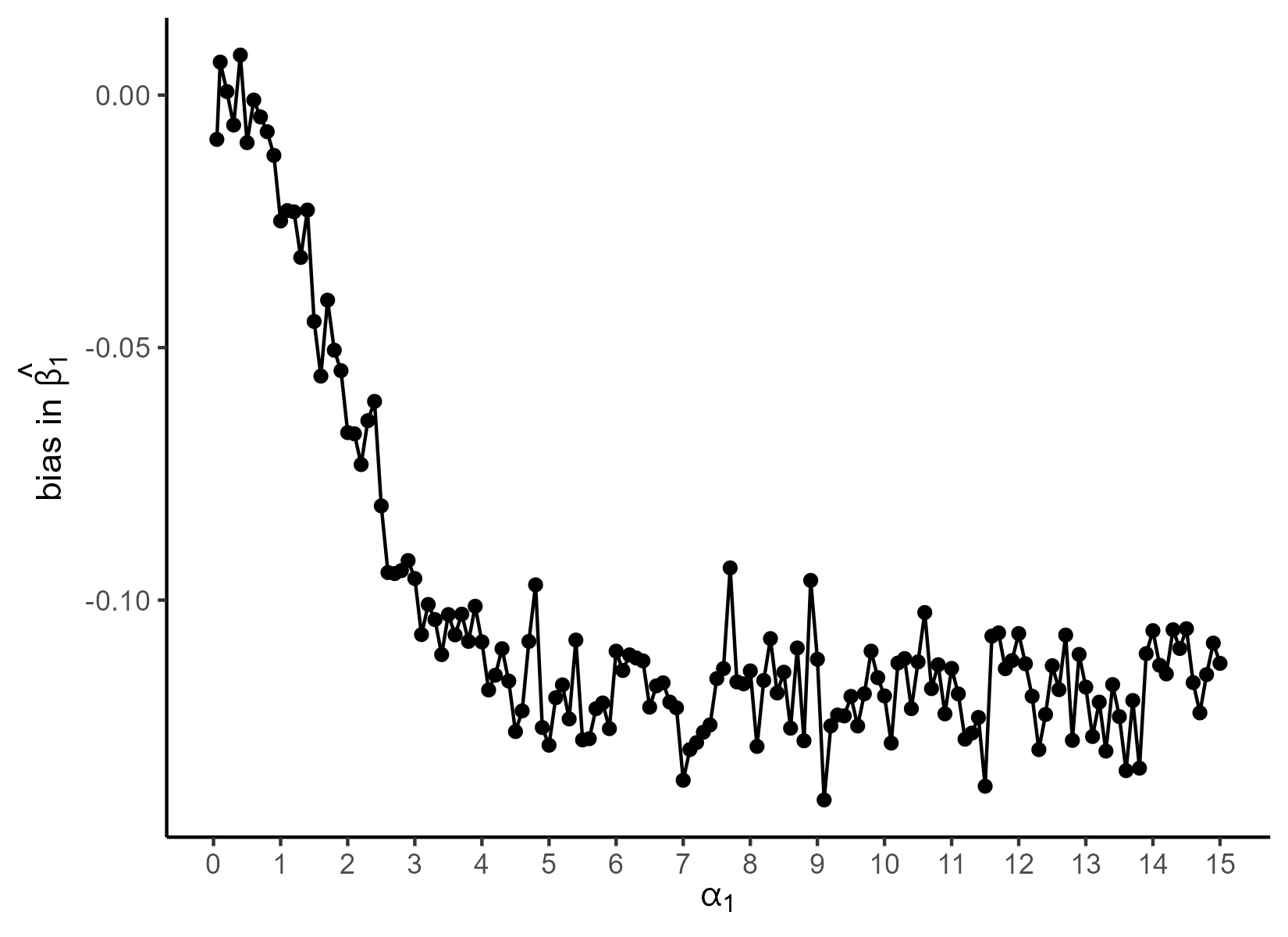}
  \caption{Trends in bias as the $\alpha_1$ is increased, holding all other parameters constant ($\sigma_{b}=\sqrt{2}$, $\sigma_{\eta}=1$, $\tau=200$, $\alpha_{0}=200/3$, $\sigma_{\epsilon}=5$, and $\gamma_{0}=-1$).} \label{fig:alpha_1plot}
  \end{center}
\end{figure}
We can see from Figure \ref{fig:alpha_1plot} that the bias generally increases in magnitude with increasing $\alpha_1$, before plateauing after the threshold at approximately -0.1 is reached, which is approximately one tenth of a standard deviation (using units in $\sigma_{b}$) below zero. This indicates a magnitude of bias that we consider non-negligible, but not excessively large. 
\subsection{Random slope for time}
\begin{figure}
\begin{center}
  \includegraphics{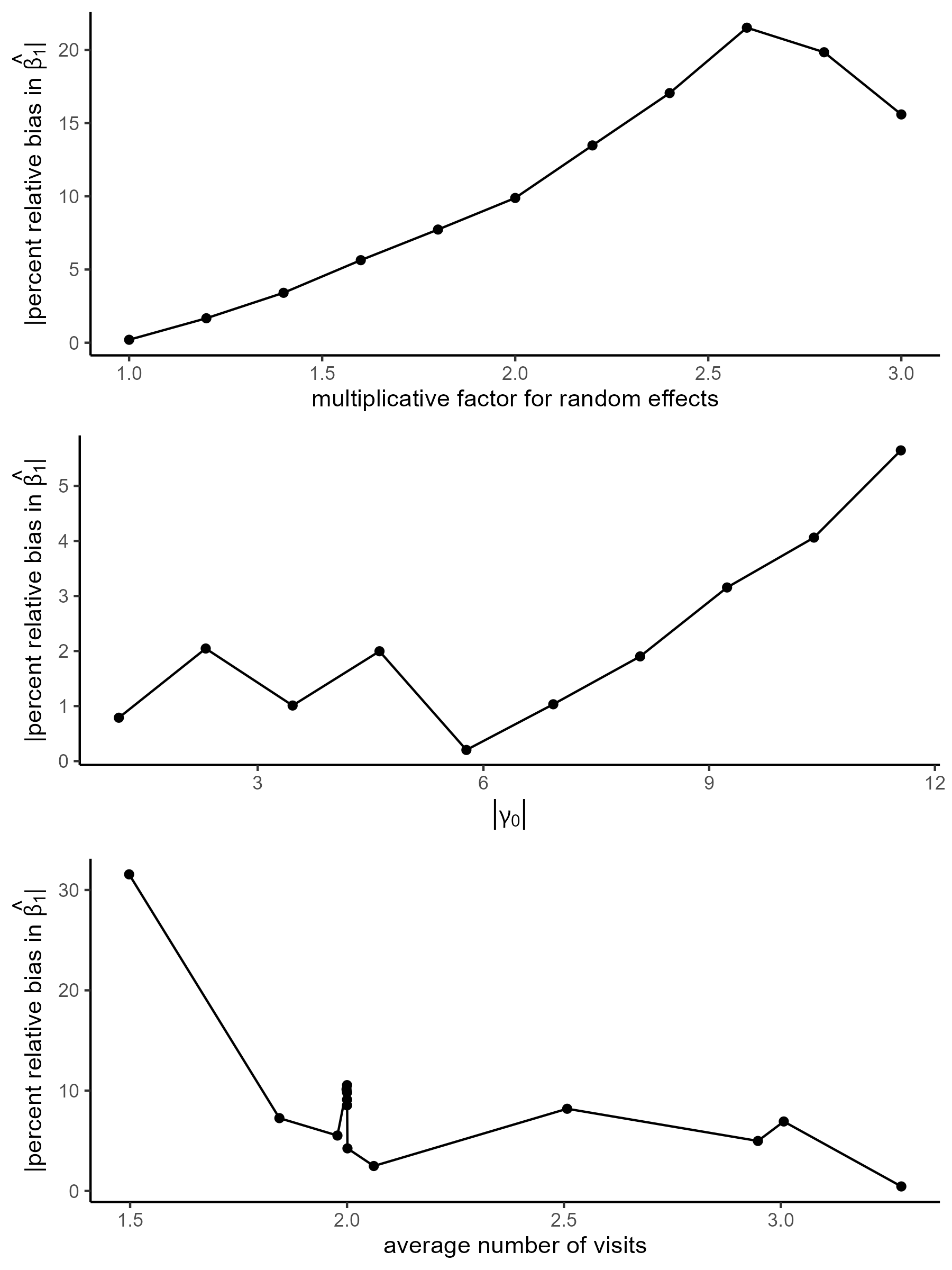}
  \caption{Trends in percent relative bias (bias/true value $\times$ 100) as a multiplicative factor for the random effects (i.e. the number by which we multiply the standard deviations of both the random intercept and random slope for time), strength of the association between the visit and outcome models is increased, and average number of visits per individual(by decreasing $\alpha_0$) is increased, respectively, holding all other parameters constant ($\sigma_{b}=\sqrt{2}$, $\sigma_{b_{1}}=
  \sqrt{2}/100$, $\rho_{b}=-0.9$, $\sigma_{\eta}=10/\sqrt{3}$, $\tau=200$, $\alpha_{0}=200/3$, $\sigma_{\epsilon}=5$, $\beta_{0}=0$, $\beta_{1}=-0.01$, and $\gamma_{0}=-10/\sqrt{3}$).} \label{fig:randomslope_theoryplot}
  \end{center}
\end{figure}

We continue with adding a random slope for time, although rather than using the complex and computationally expensive analytical expression for bias that can be found in the Appendix of this paper, we compute the bias by fitting linear mixed models on the outcome alone, using the nlme R package. The estimand of interest is the fixed effect for time. We specify the submodels as
\begin{align}
    Y_{i}(t) &= \beta_0 + \beta_{1}t  +b_{0i} + b_{1i}t + \epsilon_{i}(t) \label{Yranslope} \\
    S_{ij}&= \alpha_0 +  \gamma_{0}b_{0i} + \gamma_{0} b_{1i}t_{ij} + \eta_{ij} \label{Sranslope},
\end{align}
where the random intercept $b_{0_{i}}$ and random slope for time $b_{1i}$ are normally distributed with standard deviations $\sigma_{b}$ and $\sigma_{b_{1}}$ and correlation $\rho_{b}$. We note that $t_{ij}$ is the time of the $j^{th}$ visit for individual $i$ and again here we simulate 100,000 individuals. \\
\newline
Figure \ref{fig:randomslope_theoryplot} shows that the trends under an intercept-only model (as shown in Figure \ref{fig:fourpanelplot})  still hold true now that we have added fixed and random slopes for time. 

\subsection{Summary}
McCulloch et al \citep{mcculloch2016biased} found that estimators of parameters in the outcome model that are associated with the random effects in the visit process model will be biased, but those unconnected to the visit process random effects will be estimated with little or no bias. \\
\newline
Similar to \citep{mcculloch2016biased}, we identified under what conditions in practice the fixed intercept will be estimated with bias, but in this new context where a random intercept links the outcome to a visit process which has memory embedded into it. We found that the bias worsens with increasing variance of the random intercept, increasing strength of the association between the visit and outcome models, and decreasing average number of visits per individual. In addition, we showed that these trends held true for the bias in estimating a time effect in the presence of a random slope for time. \\
\newline
In terms of the baseline covariate effects, we found that if the covariates are unconnected to the visit process (i.e. no corresponding random effect in the outcome model; independent of $N_{i}(\tau)$), there is no bias. However, if there is a shared binary baseline covariate introducing an additional link between the visit and outcome processes, the estimator of the baseline covariate effect $\beta_{1}$ will be biased, even without having a corresponding random slope in the outcome model, which marks a difference between our results and that of McCulloch et al \citep{mcculloch2016biased}. \\
\newline
We now consider a joint modelling approach that accommodates correlation between visit and outcome models through both random effects and error terms. We first propose the joint modelling framework, and then conduct simulation studies, which will evaluate and compare the performance of the joint model to the univariate model, under more realistic parameter settings than those used in the preliminary simulations in this section.
\section{Joint model}
There are several existing examples of joint models in the literature \citep{gasparini, ryu2007, seo2020joint, zaagan2020bayesian}. These methods are distinguished mainly by their varied formulations of the visit process as different time-to-event models; Gasparini et al \citep{gasparini} model the intervals between visits assuming a Weibull baseline hazard, Ryu et al \citep{ryu2007} propose a Bayesian approach modelling the visit times using a Cox proportional hazards model, and Seo et al \citep{seo2020joint} assume an exponential distribution for the gap times with an extension to a Bayesian framework by Zaagan \citep{zaagan2020bayesian}. This variety in modelling choices occurs because there is no single widely adopted specification of the visit process, and the choice can be highly problem-specific. Moreover,  in practice it is very difficult to specify a model that accurately captures the intricacies of the often highly irregular inter-visit timings \cite{neuhaus2018analysis}. This concern around irregular and heavy-tailed visit intervals can be allayed through modelling the physician's recommendation on when the next visit should occur instead of the observed visit interval. We note that in Equation \eqref{eqn:Smodel} and the results thereafter, we assumed that patients completely adhered to the physician's recommendation, and thus, the observed interval $S_{ij}$ was simply equal to the recommendation assigned by the physician at the previous visit. This was done to simplify the derivation of theoretical results, since then we could assume that $S_{ij}$ followed a normal distribution, and although a log-normal distribution would accommodate the positive support of $S_{ij}$, this would be a poor choice due to its heavy tails. \\
\newline
The physician's recommendation on when the next visit should occur (hereafter referred to as recommended visit interval) is included in all modern EHRs. In previous work \citep{mypaper1}, we demonstrated how recommended visit intervals are useful for understanding irregularity, and so should be extracted for other analytic purposes outside of our proposed joint model. In this paper, we operate under the routine visits assumption \citep{ep2020}, which posits that the time to the next visit is based solely on the recommended interval assigned by the physician at the previous visit, and any lack of adherence to scheduled visits arises from external, non-health related factors (e.g. family vacation, physician’s work schedule), and so jointly modelling the outcome and recommended visit intervals is equivalent to modelling the outcome, recommended and observed intervals all together. \\
\newline
There are at least four advantages to using recommended rather than observed intervals in the analysis. First, they are neither heavy-tailed nor censored, so we are no longer restricted to using time-to-event approaches to model the visit process. Second, recommended intervals act as an alternative characterization of the visit process that effectively reduces the random noise seen in the observed intervals without sacrificing any important information; this provides a possible solution to the intervals between visits being highly irregular in practice and not following any easily specified data-generating mechanism \citep{neuhaus2018analysis}. Third, our proposed joint model both captures the underlying data generating mechanism more accurately than previous models and embeds memory into the visit process. Fourth, previous joint modelling frameworks \citep{gasparini, ryu2007, seo2020joint, zaagan2020bayesian} have formulated an association only through shared or correlated random effects, while our proposed joint model allows for relaxation of the simplifying assumption that there is no time-varying link between the visit and outcome processes.\\
\newline
We note that when we consider our longitudinal response as being the bivariate vector $(Y,R)^{T}$, including the disease outcome $Y$ and the recommended interval $R$, $S$ is ignorable \citep{farewell} (see Appendix for influence diagram and more detailed explanation).\\
\newline
Now, we delve into the details of specifying our proposed joint model, beginning with defining the key notation.
\subsection{Specifying the joint model with recommended visit intervals}
We note that throughout this paper, when it is useful for clarity of presentation of the disease outcome process, we use the simplified notation $Y_{ij}=Y_{i}(T_{ij})$ where $T_{ij}$ is the time of the $j^{th}$ visit for subject $i$. We let $R_{ij}$ denote the recommended visit interval for patient $i$ between visit $j$ and $j+1$, assigned by the physician at the end of visit $j$. We let $X_{ij}$ be the covariate vector for the longitudinal outcome for individual $i$, and let $G_{ij}$ be the covariate vector for the recommended interval process. Since we have conceptualized $R_{ij}$ as a function of $Y_{ij}$, we have that $X_{ij}$ and $G_{ij}$ overlap, and $G_{ij}$ may contain additional covariates beyond $Y$ that affect the recommended visit interval. We note that we do not consider any time-varying covariates other than time itself. We let $Z_{ij}$ represent the model matrix for the random effects for the disease outcome process, and $L_{ij}$ for the observation process.\\
\newline
We model the observation process and the longitudinal outcome process using a bivariate linear mixed model $(Y,R)$, with the $Y$ and $R$ sub-models being linked through both correlated random effects and residual errors. \\
\newline
We specify the disease outcome submodel as
\begin{align}
      Y_{ij} &= X_{ij}^{T}\beta + Z_{ij}^{T}b_{i} + \epsilon_{ij},
     \end{align}
where $\beta$ is the $p$-dimensional vector of fixed effects parameters for the population and $b_{i}$ is the $q$-dimensional vector of random effects for the $i^{th}$ subject.\\
\newline
The recommended visit interval submodel, with $k$-dimensional vector of random effects, is similarly specified as
\begin{align}
      R_{ij} &= G_{ij}^{T}\alpha +L_{ij}^{T} u_{i} + \zeta_{ij},
     \end{align}
     \begin{align*}
 \intertext{where}
      \begin{pmatrix}
    b_{i}\\
    u_{i}
    \end{pmatrix}
     &\stackrel{iid}{\sim} N_{q \times k} \left(\mathbf{0}, \mathbf{\Psi} \right) 
 \intertext{and}
     \begin{pmatrix}
    \epsilon_{i}\\
   \zeta_{i}
    \end{pmatrix}
     &\stackrel{iid}{\sim} N_{2N_{i}(\tau)} \left(\mathbf{0}, \mathbf{\Sigma}_{i} \right).
\end{align*}
$\mathbf{\Psi}$ is an arbitrary matrix, as this general formulation allows for all correlated random effects, independent random effects, or a mixture of both (e.g. some correlated random effects, some constrained to be independent). $\epsilon_{i}$ and $\zeta_{i}$ are vectors of the $N_{i}(\tau)$ within-subject errors for subject $i$ in the outcome and recommended interval process, respectively. The random effects are independent of the within subject errors. To ensure that $\mathbf{\Sigma}_{i}$ is positive definite, we use a separable (Kronecker product) structure.
\subsection{Separable model specification}
We draw on the literature on multivariate spatial process models \citep{spatialhandbook} and specify a separable covariance matrix for the within-subject errors as
\begin{align} \label{eqn:covstructure}
     \mathbf{\Sigma}_{i} &=  \mathbf{\Lambda} \hspace{2 pt}  \mathbf{\otimes} \hspace{2 pt} \mathbf{\Omega}_{i},
\end{align}
where $\mathbf{\Lambda}$ is a $2 \times 2$ time-invariant matrix describing the variances ($\sigma_{\epsilon}^{2}$ and $\sigma_{\zeta}^{2}$) and covariance ($\rho_{\epsilon}\sigma_{\epsilon}\sigma_{\zeta}$) in the within-subject errors among the two responses, $Y$ and $R$. $\mathbf{\Omega}_{i}=\mathbf{\Omega}_{i}(d, \mathbf{t_{i}})$ is a $n_{i} \times n_{i}$ time-dependent correlation matrix involving the parameter $d$ and the $n_{i}$ visit times $\mathbf{t_{i}}$, and $\otimes$ is the Kronecker product. This structure ensures positive definiteness, as $\mathbf{\Lambda}$ and $\mathbf{\Omega}_{i}$ must be positive definite. This structure asserts that the correlation within the two separate responses $Y$ and $R$ over time, as well as the cross-correlation between $Y$ and $R$ at different times, all share the same correlation structure, specified in the matrix $\mathbf{\Omega}_{i}$. $\mathbf{\Omega}_{i}$ can take on a wide variety of forms, as long as it is positive definite. Examples of common structures include the spatial correlation structures exponential, Gaussian, linear, rational quadratic, and spherical \citep{pinheiro}. See the Appendix for an extension incorporating a nugget effect. We also note that further modifications to this general structure are possible, such as allowing  $\mathbf{\Sigma}_{i}$ to vary by treatment group.\\
\newline
From Section 2, we know that we should expect to see bias in univariate outcome models when we have common covariates in the visit and outcome models and/or correlated random effects. We also have elucidated what factors make the bias worse in simplistic cases. In this section we have formulated a joint model that produces asymptotically unbiased estimators, under the assumptions that the visit and outcome processes are conditionally independent given the random effects, and that the model is correctly specified. What remains unknown is how large the bias in the univariate models is in realistic scenarios, nor do we know how well the joint model does at correcting this bias in small-to-moderate (100-200) sample sizes.

\section{Simulation studies}
In the following simulation studies, we fill the aforementioned gap in knowledge by setting more reasonable parameter values, guided by the dataset we will consider in section 5. We explore settings where we hypothesize we would see bias in the univariate model, based on the results in Section 3, and then we assess how the bias changes as we tone the parameters down from the more extreme end of realistic to less extreme.\\
\newline
We conduct 3 simulation studies with a range of specifications of the outcome and visit processes, and the association between the two processes.\\
\newline
The overarching goal of the simulation is to examine when a univariate model is likely to perform well, and when a joint model should be considered. We know the univariate model will be biased, but we need to discern when the bias is large enough to warrant a joint model, considering the risks of mis-specification, as noted by \citep{neuhaus2018analysis}.\\
\newline
We simulate sample sizes of 100 and 200 individuals under the first simulation set-up, and 200 individuals for the other two set-ups. We consider a study period of two years. Our theory from Section 3 identified parameter settings that exacerbate the bias in univariate linear mixed models under simplistic simulation settings. Each of our chosen simulation studies represents a realistic simulation setting with base-case parameter values chosen so as to exacerbate bias while still falling in the range of practical plausibility. These parameter values are then altered in the direction that, based on the results from Section 3, should reduce the bias. For each data-generating mechanism, we discuss the specific hypotheses we investigate, and the initial settings that we hypothesize would make the univariate approach at risk of bias.\\
\newline
We fit two models to each simulated dataset:
\begin{enumerate}
    \item Model $Y$:  the linear mixed model on the outcome alone, disregarding the observation process completely
    \item Model $(Y,R)$: the joint model of the outcome and recommended visit intervals that corresponds to the true data-generating mechanisms
\end{enumerate}
We will assess bias and empirical-based standard errors for the estimates of interest. We use 2000 replications to keep the Monte Carlo standard error for the bias at 0.0005 or lower.\\
\newline
All data simulation, summarizing of results, and the majority of modelling was done using R version 4.0.5.

\subsection{Estimation of time trends in the presence of random slopes}
This simulation study investigates whether the simplified case of random slopes we explored in Section 3 extrapolates to more complex settings, and with using more realistic parameter values. This simulation study also explores the added complexity of linking the outcome and visit processes not only through shared/correlated random effects but also through correlated error terms. We note that we did not derive a theoretical expression under correlated error terms, since it would be even more difficult to interpret than the general expression we have included under independent and identically distributed error terms.

\subsubsection*{Data-generating mechanism}
We simulate data according to our proposed joint model as
\begin{align}
    Y_{i}(t) &= \beta_{0} + \beta_{1}t +b_{0i} + b_{1i}t + \epsilon_{i}(t) \\
     R_{ij} &= \alpha_{0} + \alpha_{1} Y_{ij} +  u_{0i} +  u_{1i}t_{ij} + \zeta_{ij}.
\end{align}
 The fixed intercept of the longitudinal submodel is $\beta_{0}=7$, and the fixed effect of time is $\beta_{1}=-0.10$. The estimand of interest is the effect of time, $\beta_{1}$. The recommended intervals use the time scale of years, with the fixed intercept of the recommended interval submodel $\alpha_{0}$ set to 1, based on the logic that if there is no disease activity ($Y_{ij}=0$), the physician might assign a relatively high recommended visit interval of 1 year. The coefficient associated to the effect of the disease outcome measured at the current visit is $\alpha_{1}=(2/52-1)/15$, based on the idea if there is very high disease activity ($Y_{ij}=15$ for this simulated data), the physician might assign a short recommended interval of 2 weeks. \\
\newline
The random effects $\big(b_{0i}, u_{0i},  b_{1i},  u_{1i} \big)^{T}$ are simulated from a multivariate normal distribution with null mean vector and a variance-covariance structure with standard deviations for the longitudinal submodel $\sigma_{b_{0}}=1.6$ and $\sigma_{b_{1}}=1.2$ and $\sigma_{u_{0}}=0.06$ and $\sigma_{u_{1}}=0.05$ for the recommended interval submodel. The processes are linked through the correlation between the random intercepts $b_{0_{i}}$ and $u_{0i}$, and the correlation between the random slopes $b_{1i}$ and $u_{1i}$. We set both of these correlations to be -0.7, to induce a strong association between the outcome and visit processes. Within the longitudinal submodel, we allow for $b_{0_{i}}$ and $b_{1i}$ to be correlated, and set this correlation to be -0.5. For simplicity, all other correlations are set to zero.\\
\newline
The within-subject errors for the longitudinal submodel $\epsilon_{i}(t)$ are generated following an exponential spatial correlation structure with standard deviation $\sigma_{\epsilon}=1.5$, range $d=0.5$, and nugget $c_{0}=0.4$. The within-subject errors for the recommended visit interval model are generated as i.i.d samples from a normal distribution with standard deviation $\sigma_{\zeta}=0.05$. Note that since $R$ is simulated as a function of $Y$, the overall residual error for $R$ (i.e. in the analytic model) inherits some of the residual error from $Y$, causing the errors to be correlated.\\
\newline
The observed intervals between visits $S_{ij}$ are simulated from a Weibull distribution with shape parameter=10 and scale parameter equal to the corresponding recommended interval $R_{ij}$. This set-up produces a mean of 5.2 visits per individual.
\subsubsection*{Secondary hypotheses}
We also investigate the impact of: 1) increasing the length of the study period, 2) decreasing the absolute value of the correlation between the random slopes for time for the visit and outcome processes, and 3) decreasing the variance of all the random effects for the visit and outcome process submodels. Based on the theory from the previous section and Figure \ref{fig:randomslope_theoryplot}, the hypothesis is that the absolute value of the bias will: decrease as the length of the study period increases, as the correlation between random slopes decreases, and as the random effect variances decrease.

\subsection{Treatment effect with differential visit frequencies}
For the second simulation study, we formulate a hypothesis based on the theory extension involving the addition of a binary baseline covariate. From theory we found that increasing the strength (magnitude) of the effect of the covariate on the visit process led to greater bias in estimating the effect of the covariate on the outcome. Thus, for this simulation study we formulate a scenario with a randomly assigned binary treatment that causes patients under the treatment to visit much more frequently compared to the control group. We hypothesize that the treatment having such a strong effect on the visit process will make the univariate outcome model susceptible to incurring bias in estimating the treatment effect.
\subsubsection*{Data-generating mechanism}
\begin{figure}[ht]
\begin{center}
  \includegraphics[scale=0.7]{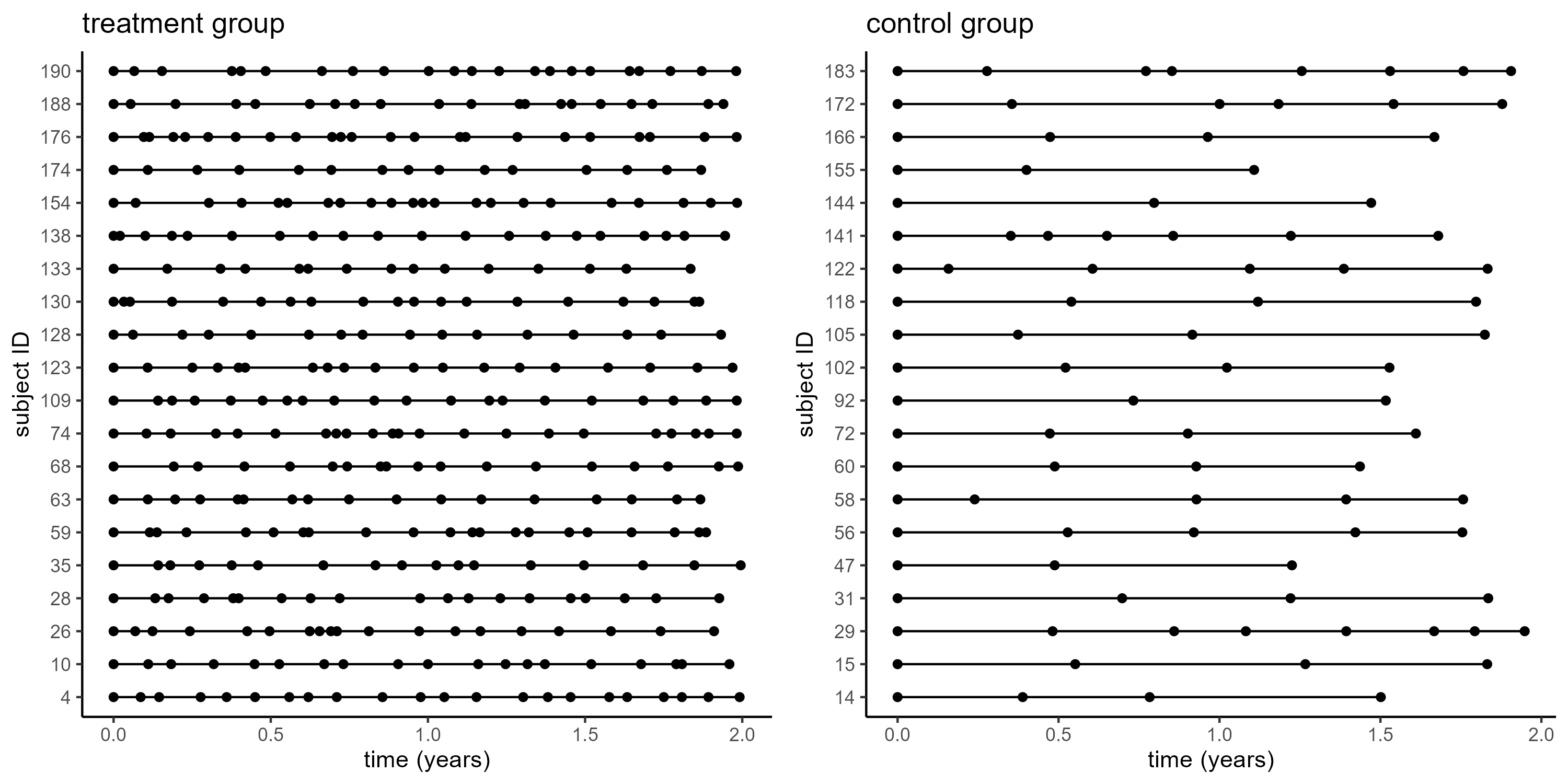}
  \caption{Simulated visit times for a random subset of 20 patients from each treatment group. Each dot represents a visit, and the solid lines indicate times where the patient was not lost to follow-up but there was no visit.} \label{fig:treatmentplot}
  \end{center}
\end{figure}

We use a generally similar set-up to study \# 1 in terms of the fixed and random effect parameter values, with the addition of the treatment variable. For simplicity, we also restrict the within-subject errors in the $Y$ submodel to be i.i.d rather than autocorrelated. We note that this is a simplified joint model compared to the one we proposed in the previous section. That is, in this set-up, we have residual variation but no temporal structure and this can be be formulated by constraining $\Omega_{i}$ in (\ref{eqn:covstructure}) to be the identity matrix. Thus, the data are generated according to the following joint model
\begin{align}
     Y_{i}(t) &=  \beta_{0} + \beta_{1}t + \beta_{2}X_{i} +b_{0i} + b_{1i}t + \epsilon_{i}(t) \\
 R_{ij} &=
    \begin{cases}
      \alpha_{0}  + \alpha_{1}Y_{ij} +   u_{0i} + u_{1i}t_{ij} + \zeta_{ij} & \text{if $X_{i}=0$}\\
      \alpha_{0} + \alpha_{2} + \zeta_{ij} & \text{if $X_{i}=1$},\\
    \end{cases}       
\end{align}
where $X_{i} \sim \text{Bernoulli}(0.5)$, $\beta_0=7$, $\beta_{1}=-0.1$, $\beta_{2}=-1$, $\alpha_{0}=1$, $\alpha_{1}=(2/52-1)/15$, $\alpha_{2}=-46/52$. The random effects and $S_{ij}$ are generated in the same manner as in simulation study \#1. The error variances are also kept constant at $\sigma_{\epsilon}=1.5$ and $\sigma_{\zeta}=0.05$. \\
\newline
The mean number of visits per person was 18.7 under the treatment and 4.8 under the control group. \\
\newline
The logic behind this set-up is as follows: at the first visit, patients are assigned to either the treatment ($X_{i}=1$) or control ($X_{i}=0$) group, with equal probability. They remain in their assigned group for the entirety of the follow-up period. The physician recommends that subjects in the treatment group come back to visit the clinic every 6 weeks, regardless of their disease condition. The recommended interval for subjects in the control group is determined as in simulation study \#1. Figure \ref{fig:treatmentplot} shows how this set-up creates a treatment group of patients who consistently visit very frequently, and a control group who display typical, irregular visit patterns.\\

\subsubsection*{Secondary hypothesis}
To continue investigating trends in bias in set-up 2, we also study less extreme differences in the visit process between the control and treatment groups. That is, we assess the case where all patients follow the control group visiting scheme. Based on the theory presented earlier, the hypothesis is that the bias in the estimate of the treatment effect in the univariate outcome model should decrease.

\subsection{Restricting estimation to early visits}
For this simulation study, we formulate a hypothesis based on the theory that fewer visits per person lead to higher bias. We generate data using an exponential decay function of time so that the parameter of interest is estimated only using earlier time points (when not many visits have occurred yet). We hypothesize that the linear effect of time will be biased in this context.
\subsubsection*{Data-generating mechanism}
In this study, we omit the random slope for time that we had in the previous set-ups, for simplicity. We have
\begin{align}
    Y_{i}(t) &= (\beta_0 + \beta_1 t)w(t) + \beta_3(1-w(t)) +b_{0i} + \epsilon_{i}(t)\\
    R_{ij} &= \alpha_0 + \alpha_{1}Y_{ij} + u_{0i} + \zeta_{ij},
\end{align}
where $\beta_{0}=7, \beta_{1}=-5, \beta_{2}=2$, $w(t)=e^{-4t}$, and for the parameters that were also used in the previous simulation studies, we used similar or identical values: $\alpha_0=1$, $\alpha_1=(2/52-1)/12$, $\sigma_{\epsilon}=1.2$, $\sigma_{\zeta}=0.05$, $\sigma_{b_{0}}=1$, $\sigma_{u_{0}}=0.06$, and with a correlation of -0.7 between the random intercepts. $S_{ij}$ is generated in the same way as in the previous simulations. \\
\newline
The average number of visits per person under this set-up was 3.7.
\subsubsection*{Secondary hypothesis}
We also generated datasets with $w(t)=e^{-2t}$. The hypothesis is that we as slow the rate of decay, we are able to use more of the observations to estimate $\beta_{1}$, so we expect that the bias will decrease.

\subsection{Results}
For estimating the time effect in Simulation Study \#1, there was 31\% relative bias in the univariate model when all three bias-causing factors were at the high level. This bias decreased as any one factor was shifted to the ``medium" level, and there was $<$ 15\% bias for the joint model across all scenarios (see Table \ref{table:sim1}).\\
\newline
We note that under Simulation Study \#1, we also assessed the performance of several other modelling approaches. A trivariate model $(Y,R,S)$ of the outcome, recommended visit intervals, and observed visit intervals formulated according to the true data-generating mechanism (see Appendix for specification) performed well, with relative bias of 2.9\% with ESE 2.7. A bivariate model of the outcome and observed visit intervals, $(Y,S)$, using a framework similar to \citep{gasparini} had a relative bias of 16\% with ESE 2.6. A generalized estimating equation (GEE) ignoring the visit process had relative bias of -750\% with ESE 6.2, whereas using inverse-intensity weighted (IIW-GEE) produced a relative bias of 36\% with ESE 12.3. See the Appendix for a descriptive table on the visit process characteristics under the various scenarios and for results using a sample size of 100 individuals.\\
\newline
Similarly, for estimating the parameters of interest in Simulation Studies \#2 and 3 there was 3 and 6\% relative bias, respectively, in the univariate model when the scenario-specific bias-causing factor was at the high level, and this decreased when the given factor was toned down to the low level. There was $<$ 0.5\% relative bias for the joint model across all scenarios (see Table \ref{table:sim2}).

\section{Application}
\label{sec:application-section}
We illustrate our approach by analyzing data from a clinic-based cohort of patients diagnosed with juvenile dermatomyositis (JDM).

\subsection{The JDM dataset}
JDM is a rare autoimmune disease in children, with common symptoms of rashes and muscle inflammation that can cause difficulties with walking and other day-to-day tasks. JDM is chronic and thus, understanding disease trajectory over time is important.\\
\newline
This study was approved by the Hospital for Sick Children (SickKids) Research Ethics Board (REB \#1000019708). The study population consisted of patients enrolled in the specialized JDM clinic at SickKids who visited the clinic at least twice between June 1, 2000 and May 31, 2018, and had a documented date of diagnosis in their chart. The final sample size was 149 patients with a combined total of 2912 visits. \\
\newline
Patients were followed up on an as-needed basis. Follow-up terminated upon transitioning to adult care, or upon transferring to another paediatric care provider. Follow-up was also censored at the end of the study period for some individuals who were still actively enrolled in the clinic.\\
\newline
Data were collected from patient charts within the electronic health record database. General clinic letters and physician notes were used to extract physician recommendations on when to next visit the clinic, and comments regarding follow-up times were also extracted to provide additional information, such as explanations as to why some recommended follow-up times were missing. The physician typically recommends that sicker patients visit the clinic more frequently. Thus, the visits are not only irregularly spaced, but are also more frequent when patients are sicker.\\
\newline
This JDM clinic generally runs every two weeks, so the standard recommended intervals are multiples of two weeks and the typical shortest possible recommended interval would be two weeks, but this is not strictly adhered to, as patients can come in sooner if more urgent care is needed. In addition, at the very beginning of starting treatment, patients are seen 3 or 4 times at 2 weekly intervals to monitor for treatment side effects. When recommended intervals were missing, the reason for missingness was extracted whenever possible. For example, the next visit may be “pending family decision”, “pending MRI”, or “pending consultation with another doctor”. 
\subsection{Modelling objectives}
The outcome of interest is a modified disease activity score (DASmod) \citep{dasmod}. The DASmod ranges from 0 to 12, and a higher score indicates more severe disease activity. The target of inference in this study was the disease trajectory over time, where time is measured in years since diagnosis. Bayesian models were fit using Stan (rstan Version 2.21.2). All other analyses were conducted using R version 4.1.0.\\
\newline
We also use several diagnostic methods to examine whether a joint model would be useful: 1) calculate descriptive statistics of the number of visits per patient, 2) fit a univariate model on the outcome and compute the variance of the random effects in relation to the residual variance, and 3) compute predicted subject-specific random effects from the model in \#2. Then, we also fit a univariate model on the recommended visit intervals and computed predicted subject-specific random effects. We then produced a scatterplot of the random effects from the outcome model versus the random effects from the recommended visit interval model. \\
\newline
We used our proposed framework to jointly model DASmod and the recommended visit intervals. Note that we performed a Box-Cox transformation on the recommended visit intervals to better meet the assumption of normally distributed errors, and so the notation $R$ in this section refers to Box-Cox transformed versions of the original $R$. We used the model
\begin{align}
    Y_{ij} &= \beta_{0} + \frac{\beta_{1}}{{(1+T_{ij})}^{2}} + \beta_{2}\frac{\log(1+T_{ij})}{{(1+T_{ij})}^{2}} +b_{i} + \epsilon_{ij}\\
    R_{ij} &= \alpha_{0} + \frac{\alpha_{1}}{{(1+T_{ij})}^{2}} + \alpha_{2}\frac{\log(1+T_{ij})}{{(1+T_{ij})}^{2}}  + u_{i} + \zeta_{ij},
\end{align}
where the random intercepts $b_{i}$ and $u_{i}$ are normally distributed and correlated, and the error terms $ \epsilon_{ij}$ and $\zeta_{ij}$ are normally distributed with the separable covariance matrix $\mathbf{\Sigma}$ described in the joint model section, using an exponential spatial correlation structure for both $Y$ and $R$ processes. We used a fractional polynomial function of the time since diagnosis as the fixed effects portion, following \citep{ep2016}.
\subsection{Results}
\begin{figure}
    \centering
    \includegraphics[width=170mm]{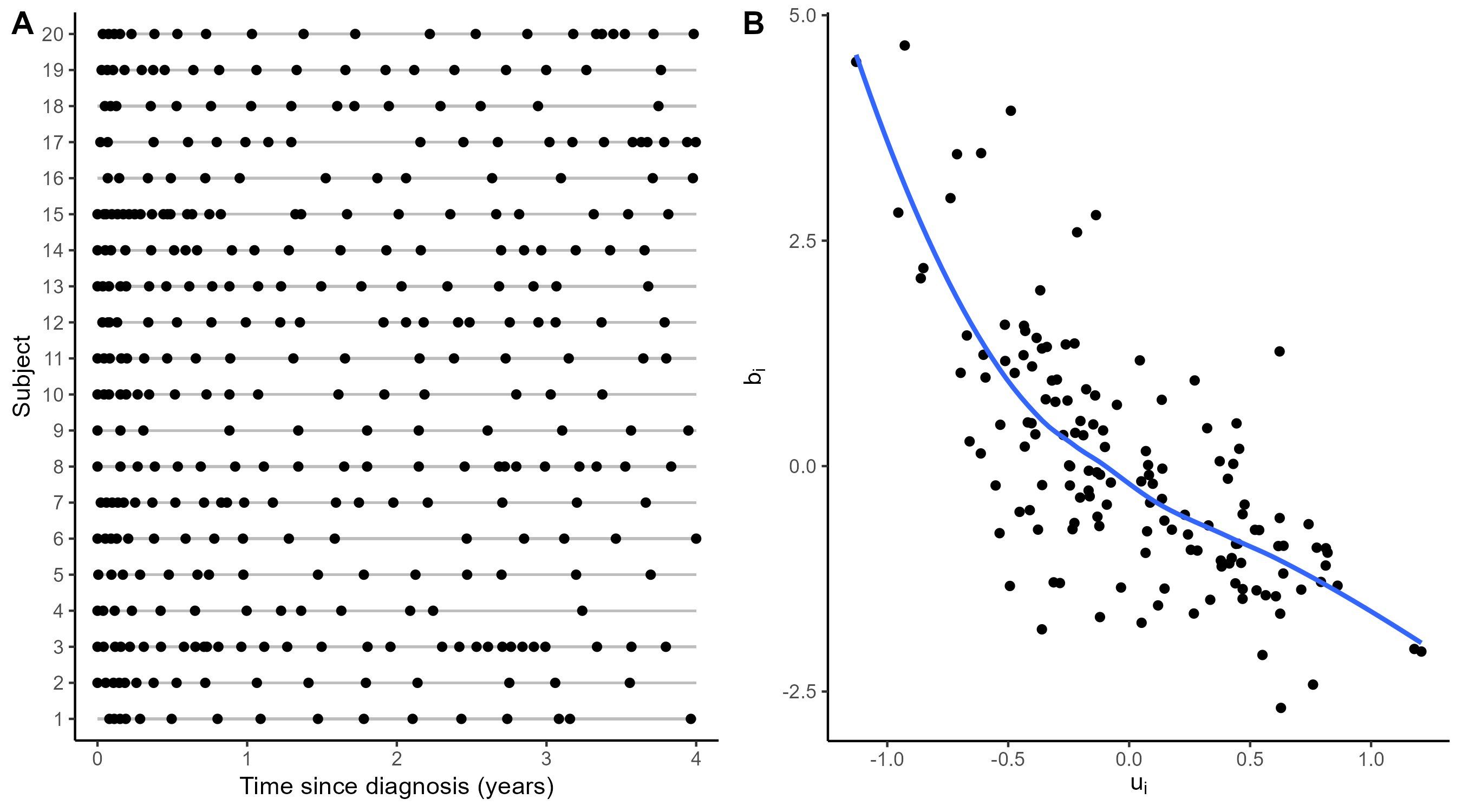}
    \caption{A: Visit times for a random subset of 20 patients in the JDM dataset over their first 4 years of follow-up since being diagnosed. Each dot represents a visit, and the grey solid lines indicate times where the patient was still being followed at the clinic, but there was no visit. B: plot of estimated random intercepts from $Y$ univariate linear mixed model and $R$ univariate mixed model.}
    \label{fig:JDM_diag}
\end{figure}

\begin{figure}[ht]
  
  \includegraphics[width=170mm]{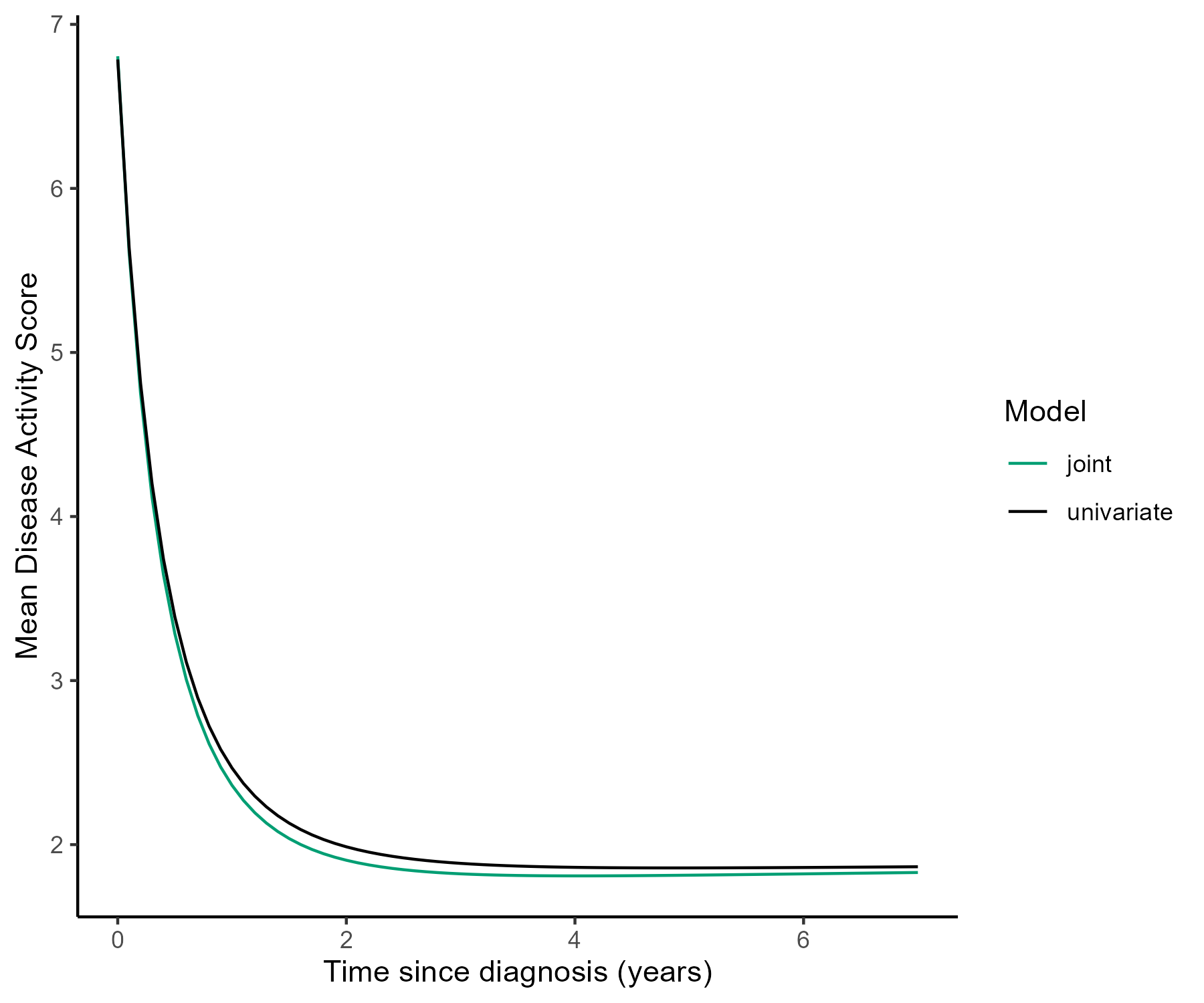}
  \caption{Fitted expected disease activity score as a function over time using our proposed joint model and a univariate model for the disease outcome that ignores the informative visit process.}\label{fig:jm}
 \end{figure}

 There were 149 patients enrolled in the clinic after diagnosis, and 84 (56\%) were followed up until transitioning to adult care. The remaining patients either had their follow-up censored at the end of the study period, or transferred to another hospital, or follow-up was stopped for some other reason before the patient reached age 18. The median duration of follow-up was 6.9 years (interquartile range 3.4-10.9 years). The median recommended interval was 3 months (interquartile range 1.4-4 months), ranging from 3 days to 14 months. The median observed interval between visit was 2.8 months (interquartile range 1.4-4.3 months), ranging from 1 day to 3.4 years. Figure \ref{fig:JDM_diag}A depicts visit times from a randomly selected subset of the patients in the dataset; the gaps between visits vary both between and within patients. \\
 \newline
 For the first diagnostic for the potential for bias in a univariate mixed model, the median number of visits per patient was 18 (interquartile range 10-26). \\
 \newline
For the second diagnostic, we found that the linear mixed model on the outcome produced an estimated standard deviation of 1.36 for the random intercept and 1.69 for the residual error, and thus, an ICC of 0.39.  \\
\newline
For the third diagnostic, Figure \ref{fig:JDM_diag}B shows clear evidence of a strong correlation between the random effects from the univariate outcome and recommended visit interval models. \\
\newline
In conclusion, while the JDM dataset contains a high number of visits per patient (and we previously showed bias decreases with increasing number of visits), the other two diagnostics suggest that the univariate outcome model could be at risk of producing biased estimates. Thus, it is worthwhile to consider fitting a joint model.\\
\newline
Before fitting the joint model, we checked its underlying assumptions. The separability assumption was reasonably met, as the 95\% credible intervals for the range parameter for the univariate disease outcome model, the univariate recommended intervals model, and the joint model were overlapping (see Appendix for results).\\
\newline
Estimates of the parameters in the joint model quantifying the strength of the association between the outcome and visit processes are shown in Table \ref{table:correlationsJM}. The high magnitudes of the random intercept correlation, as well as the spatial residual correlation suggest non-ignorable dependence of the outcome process on the irregular visit process.\\
\newline
In Figure \ref{fig:jm}, we compared the univariate mixed model results to the joint model results, and we found that the estimated trajectories were very similar, with overlapping confidence intervals. Thus, this is a real-world example of a case where the univariate model performs just as well as the joint model. This may be driven by the high median number of visits per patient (18, which is at the high end of what we typically see), but there were also opposing factors that we would expect to possibly lead to bias; that is, the correlation of -0.8 between the random intercepts for the $Y$ and $R$ submodels was high and as were the intraclass correlation coefficients (ICCs) at 0.38 for $Y$ and 0.34 for $R$.

\section{Discussion}
In this paper, we have explored parametric approaches to modelling irregular longitudinal data, specifically the notable robustness of univariate linear mixed models to violations of the modelling assumptions. Mathematical derivations show that the univariate model is biased, however through simulations we found that the extent of the bias is often small and the univariate model performs similarly to the joint model. In more detail, we have three key findings regarding the trends in bias under the univariate model:\\
\newline
First, there was no bias in regression coefficients of baseline covariates that are unconnected to the visit process and have no corresponding random effect in the outcome model.\\
\newline
Second, for binary baseline covariates that are common to both the outcome and visit models, there was bias, and the magnitude of the bias increases as the effect of the baseline covariate on the visit interval size gets larger.\\
\newline
Third, three major factors led to higher bias in estimating either the intercept or the slope for time: high variance of random effects, high correlation between outcome and visit processes (typically through strongly correlated random effects), and small number of visits per individual. We discuss diagnostics for how to assess for evidence/warning signs that the univariate mixed model is at high risk of producing biased estimates in Table \ref{tab:diagnositics}, and we illustrated this process with the JDM data analysis. \\
\newline
In cases where there was bias, we found that our proposed joint model was able to eliminate the bias. We note that any kind of time-varying covariates (other than time itself) are typically endogenous, and thus require an IIW-GEE. \\
\newline
 Previous work \citep{mcculloch2016biased, neuhaus2018analysis,neuhaus2020} assessed the bias of the univariate model assuming a memoryless visit process, whereas we investigate the problem under a different visit paradigm with memory embedded into it, reflecting EHR data. We attained generally parallel results in this new visit process context to \citep{mcculloch2016biased,neuhaus2018analysis} in terms of the effect of shared/correlated random effects linking the outcome and visit processes, and the effect of varying the strength of this association. Our results differed in that we found that if there was binary baseline covariate common to both the outcome and visit processes, there was still bias even without an associated random effect. In addition, we expanded upon the existing work in the literature by examining other trends, including the effect of varying the average number of visits per patient, and the variances of the random effects (while holding the residual variance constant). \\
\newline
 In addition, the presence of some regular visits was previously observed to greatly reduce even the small bias in results produced by mixed models under irregular visit processes \citep{neuhaus2018analysis}. We note that although regular visits may not be typically found in the context of EHR data, research cohorts sometimes may instate a decision to take measurements with minimum frequency; for example, taking a urine sample at least once a year to monitor kidney function. We have not studied the impact of this on bias, but we conjecture that in the memory-embedded visit process context, this too would reduce bias. \\
 \newline
We note that while our proposed joint model requires that the analyst has access to the recommended visit intervals, in modern EHRs, such as Epic, this information is often included and could be extracted automatically. Thus, in studies going forward the proposed methods in this paper should be highly feasible. However, we note that if the recommended visit intervals are not available (e.g. if using older EHR data), approaches from the literature (e.g. \citep{gasparini}) of modelling the outcome and observed intervals are an option. \\
\newline
We also note that we did not consider stochastic time-varying covariates in this paper because they are typically endogenous, and thus require an IIW-GEE.\\
\newline
 Furthermore, our models assume that the outcome and observed intervals between visits are independent, given the recommended visit intervals (routine visits assumption), which may not be realistic in practice. Our previous work \citep{mypaper1} proposed an approach for sensitivity analysis for GEEs in this context, but there is no currently existing framework for parametric approaches (linear mixed models), and thus, this is an area for further work.\\
 \newline
In this paper we have worked with linear models only. Future work could consider binary or count outcomes, which have been previously investigated under the restriction of memoryless visit processes \citep{neuhaus2018analysis,neuhaus2020}. Future studies could also look further into the effects of model misspecification, such as regarding the shape and correlation structure of the error distribution in linear mixed effects models or the shape of the random effects distribution, building upon \citep{neuhaus2020} (they found little effect of misspeecification). Further work could also look at the interplay between random effects and correlation structures on the error terms under an irregular visit process with memory embedded (e.g. if we add additional random effects, do we still need the complex correlation structure?). Moreover, we have also modelled the recommended visit intervals using a linear model, and in some applications, for example, when there are a small number of values that recommended visit intervals can take on, it might be advisable to use a categorical distribution.\\
\newline
In summary, there is a common perception that mixed models can handle variable number and timings of visits per patient. The truth is more nuanced; there is no bias if the irregular visit times are non-informative. The contribution of this paper is to show that when visit processes have memory, as in the context of EHR data, univariate mixed models fit by maximum likelihood can be robust to informative visit processes. In addition, we have suggested both diagnostics for when joint modelling may be needed, and a joint model that accounts for the informative visit process. Thus, while univariate mixed models may be appropriate, they should not be used as a default analytic approach, but rather used only after assessing whether joint models are needed.


\bibliography{main}
\newpage

\section*{Tables}
\begin{table}[ht]
\begin{threeparttable}
\centering
\caption{Percent relative bias (ESE) in the slope for time in simulation study 1 under high, medium, and low bias settings, with 200 individuals per simulated dataset}
\begin{tabular}{@{}llllllll@{}}
\toprule
&  \multicolumn{3}{c}{Univariate} & & \multicolumn{3}{c}{$\text{Joint}$}\\
\cmidrule{2-4} \cmidrule{6-8} 
Bias-causing factor & High & Medium & Low && High & Medium & Low \\ \midrule
$\tau$ (in years) & \multirowcell{3}[0pt][l]{ -30.70 (2.7)}   & -15.41 (2.2) & \phantom{1}-8.03 (2.1) &&  \multirowcell{3}[0pt][l]{ 6.34 (2.6)} & \phantom{-1}5.52 (2.2) & \phantom{1}-2.97 (2.1)\\
$\text{Corr}(b_{1i}, u_{1i})$  &  & -20.98 (2.8) & -10.04 (2.7) &&  & -10.98 (2.8) & -13.24 (2.7) \\
Var of ran. effects  & & -15.21 (2.2) & \phantom{1}-7.13 (1.9) &&  & \phantom{1}-0.04 (2.2) & \phantom{1}-0.89 (1.9) \\
\bottomrule

\end{tabular}\label{table:sim1}
\begin{tablenotes}
      \small
      \item Percent relative bias= bias/truth $\times$ 100. ESE =empirical standard error.  For $\tau$ (length of study period in years): High=2, Medium=3, and Low=4. For $\text{Corr}(b_{1i}, u_{1i})$ (correlation between random slopes): High=-0.7, Medium=-0.3, and Low=0. For Var of ran. effects (variance of random effects): High= original setting, Medium= original variance of all random effects divided by 4, and Low= variance of all random effects divided by 10.
    \end{tablenotes}
  \end{threeparttable}
\end{table}

\begin{table}[ht]
\centering
\begin{threeparttable}

\caption{Percent relative bias (ESE) in parameters of interest in simulation studies \# 2 and \#3 under high and low bias settings, with 200 individuals per simulated dataset.}
\begin{tabular}{@{}llllll@{}}
\toprule
&  \multicolumn{2}{c}{Univariate} & & \multicolumn{2}{c}{$\text{Joint}$}\\
\cmidrule{2-3} \cmidrule{5-6} 
& High  & Low && High &  Low \\ \midrule
Simulation \# 2 & 3.378 $\hspace{1 pt} (0.47)$ & -0.801 $\hspace{1 pt} (0.49)$ $\textcolor{white}{\times 10^{-7}}$   && 0.016 $\hspace{1 pt} (0.47)$ & 0.103 $\hspace{1 pt} (0.47)$ \\
Simulation \# 3 & 5.731 $\hspace{1 pt} (0.64)$ & -1.648 $\hspace{1 pt} (0.33)$  && 0.391 $\hspace{1 pt} (0.63)$  &  $0.000$ $\hspace{1 pt} (0.33)$   \\
\bottomrule

\end{tabular}\label{table:sim2}
\begin{tablenotes}
      \small
      \item Percent relative bias= bias/truth $\times$ 100. ESE =empirical standard error. For Simulation 2, High= original set-up, and Low= homogenized version of visit process. For simulation 3, High= original set-up, and Low= decreasing rate of decay to $w(t)=e^{-2t}$. We note that for simulation study \# 2, the parameter of interest for which we report the bias is treatment effect $\beta_{2}$, and for study \#3, it is the slope for time $\beta_{1}$. 
    \end{tablenotes}
  \end{threeparttable}
\end{table} 

\begin{longtable}[]{@{}lccc@{}}
\caption{Correlation and standard deviation estimates from the joint model}\\ 
\toprule
parameter &  posterior mean & posterior sd & 95\% credible interval \tabularnewline
\midrule
\endhead
$\rho_b$: random intercept correlation & -0.82 & 0.045 & [-0.90, -0.73]\tabularnewline
$\rho_{w}$ : spatial residual correlation & -0.49 & 0.043 & [-0.57, -0.40] \tabularnewline
$\rho_{e}$ : measurement error residual correlation & 0.069 & 0.045 & [-0.15, 0.022] \tabularnewline
$\sigma_{b}$ : $Y$ random intercept st dev & 1.34 & 0.10 & [1.15, 1.55] \tabularnewline
$\sigma_{u}$ : $R$ random intercept st dev & 0.45 & 0.038 & [0.38, 0.53] \tabularnewline
$\sigma_{\epsilon}$ : $Y$ error residual st dev & 1.71 & 0.034 & [1.65, 1.78] \tabularnewline
$\sigma_{\zeta}$ : $R$ error residual st dev & 0.62 & 0.013 & [0.60, 0.65] \tabularnewline
\bottomrule
\label{table:correlationsJM}
\end{longtable}

\begin{table}
\caption{Diagnostics for how to assess for evidence or warning signs that the univariate mixed model is at high risk of producing biased estimates.}
    \begin{tblr}{
      colspec = {*{3}{X}},
      cells={
      valign=m},
      stretch=2,
      vlines
    }
   \toprule
    Bias-Causing Factor & Statistical Procedure & Bias Criterion\\ \hline
 Small number of visits per individual & Descriptive statistics  & Assess mean number of visits \\
High variance of random effects in relation to residual variance & Univariate mixed model for the outcome &   variance of the random effects divided by the residual variance \\
High correlation between outcome and visit processes &  Recommended visit intervals available: fit univariate mixed model to them. Otherwise: fit a frailty model to the observed visit intervals & Scatterplot of outcome model random effects vs. random effects/frailty terms from the univariate visit interval model\\
Covariate affects both outcomes and visit intervals & Include the covariate as a predictor in the univariate visit process model & Check for evidence of an association between the covariate and the visit intervals\\
\bottomrule
    \end{tblr}
        \label{tab:diagnositics}
\end{table}

\end{document}